\documentclass[%
 reprint,
superscriptaddress,
 amsmath,amssymb,
 aps,
 prl,
floatfix,
]{revtex4-2}

\usepackage{booktabs}
\usepackage{afterpage}
\usepackage[T1]{fontenc}
\usepackage{mathtools}
\usepackage{amssymb}
\usepackage{amsthm}
\usepackage{cancel}
\usepackage{physics}
\usepackage{multirow}
\usepackage{placeins}
\usepackage{graphicx}
\usepackage{dcolumn}
\usepackage{bm}

\newcommand{\f}{\frac}

\newcommand{\gam}{\gamma}

\newcommand{\te}{\text}

\newcommand{\appFormFactor}{A}
\newcommand{\appUncertainty}{B}
\newcommand{\appScanSpeed}{C}

\begin{document}
\raggedbottom

\preprint{APS/123-QED}

\title{Axion dark matter search with a photonic bandgap cavity haloscope\\and dielectric tuning rod over 10.25-10.45 GHz}

\author{Morgan Lynn}
 \affiliation{Department of Physics, University of Chicago, Chicago, Illinois 60637, USA}
 \author{Ankur Agrawal}
 \affiliation{Department of Physics, University of Chicago, Chicago, Illinois 60637, USA}
 \affiliation{James Franck Institute, University of Chicago, Chicago, Illinois 60637, USA}
 
\author{Arjun Ghosh}
 \affiliation{Department of Astronomy and Astrophysics, University of Chicago, Chicago, Illinois 60637, USA}
 
 \author{Sara Sussman}
 \affiliation{Fermi National Accelerator Laboratory, Batavia, Illinois 60510, USA}

\author{Steven G. Johnson}
\affiliation{Department of Mathematics, Massachusetts Institute of Technology, Cambridge, MA 02139, USA}

\author{David I. Schuster}
\affiliation{Department of Physics, University of Chicago, Chicago, Illinois 60637, USA}
 \affiliation{James Franck Institute, University of Chicago, Chicago, Illinois 60637, USA}
 \affiliation{Pritzker School of Molecular Engineering, University of Chicago, Chicago, Illinois 60637, USA}
 \affiliation{Department of Physics and Applied Physics, Stanford University, Stanford CA, 94305}

\author{Aaron S. Chou}
\affiliation{Fermi National Accelerator Laboratory, Batavia, Illinois 60510, USA}
\affiliation{Department of Physics, University of Chicago, Chicago, Illinois 60637, USA}

\date{\today}
\begin{abstract}
We report the development of a new widely tunable cavity and demonstrate its use in a search for dark matter axions. We achieve unloaded quality factors above $10^{5}$, roughly $25\times$ larger than a bare copper cavity at the same frequency, using concentric sapphire shells to reduce Ohmic losses on the cavity barrel. A rotating sapphire rod tunes our cavity mode over the $10.1-11.7$ GHz range, approximately $16\%$ of its resonant frequency. Using an amplified receiver chain, we demonstrate sensitivity to new axion parameter space by tuning the cavity over the $200$ MHz range between $10.25-10.45$ GHz ($42.4 - 43.2\, \mu \te{eV}$) to constrain the axion-to-photon coupling to $|g_{a \gam \gam}| \leq 1 \times 10^{-12}\, \te{GeV}^{-1}$. This cavity can scan its tuning range about $9$ times faster compared to a bare copper cavity when paired with a photon counting device, laying the groundwork for a definitive search for the QCD axion over $10.1-11.7$ GHz.
\end{abstract}

\maketitle


\section{\label{sec:Intro}Introduction}
\indent One of the largest mysteries in physics is the nature of dark matter. Multiple experiments point to the existence of what we call dark matter but otherwise reveal little about its intrinsic properties \cite{tanabashiReviewParticlePhysics2018, rubinRotationalProperties211980}. One of the leading dark matter candidates is the axion~\cite{weinbergNewLightBoson1978, wilczekProblemStrongInvariance1978} which also solves the strong CP problem, an unexplained symmetry in particle physics~\cite{pecceiCPConservationPresence1977, pecceiConstraintsImposedCP1977}. Axion dark matter is predicted to be composed of low mass bosons that, due to their high galactic phase space density, behave as a coherent wave with large occupation number. Experiments aiming to detect signatures of this coherent oscillation are well-established, with the most prominent being the cavity axion haloscope \cite{sikivieExperimentalTestsInvisible1983, sikivieDetectionRatesInvisible1985,duSearchInvisibleAxion2018, zhongResultsPhase12018, yiAxionDarkMatter2023}. Given that the axion's mass is not known, these cavities need to be widely tunable, with $20\%$ tunability a target for some next-generation searches \cite{kuoMaximizingQuantumEnhancement2025}. \\
\indent In this paper, we report the operation of a new cavity haloscope utilizing concentric sapphire shells following the design pioneered in Ref.~\cite{divora2022highqmicrowavedielectricresonator} and a sapphire tuning rod to achieve $16\%$ tunability over the $10.1-11.7\, \mathrm{GHz}$ frequency range. We will also discuss the cryogenic engineering required to operate such a cavity. To demonstrate this cavity's use as an axion haloscope, we conduct a search in the $42.4-43.2\, \mu\mathrm{eV}$ mass range and provide new limits on the axion-photon coupling to the $\left| g_{a \gam \gam} \right| \leq 1 \times 10^{-12}\, \mathrm{GeV}^{-1}$ level. \\
\indent Traditional haloscopes have two primary figures of merit: the power deposited by the axion in the form of a photon, and the scan rate, which describes how quickly a given haloscope can scan a range of frequencies. The axion power can be written in natural units as in Ref.~\cite{brubakerFirstResultsMicrowave2017},
\begin{equation}\label{Axion Power}
    P_{\te{ax}} = \left(g_{\gam}\f{\alpha}{\pi}\theta \right)^{2}\f{\beta}{\left(1 + \beta\right)^{2}} B^{2} CVQ_{0} m_{a}\f{1}{1+ \left(\f{2Q_{L}\Delta f_{c}}{f_c}\right)^{2}}
\end{equation}
where the first set of parentheses holds the theory parameters: $\alpha$ is the fine structure constant, the axion wave amplitude $\theta = \sqrt{2\rho_{a} / \Lambda_{\te{QCD}}^{4}}$ for local dark matter density $\rho_{a} = 0.45 \, \mathrm{GeV}/\mathrm{cm}^{3}$ \cite{readLocalDarkMatter2014} and QCD scale parameter $\Lambda_{\te{QCD}} = 77 \, \mathrm{MeV}$, and $g_{\gam}$ is a model-dependent dimensionless coupling. 
The two benchmark models are denoted KSVZ \cite{kimWeakInteractionSingletStrong1979,shifmanCanConfinementEnsure1980} and DFSZ \cite{dineSimpleSolutionStrong1981} that set $g_{\gam} = -0.97$ and $g_{\gam} = 0.36$ respectively. 
The physical coupling that appears in the full axion Lagrangian is $g_{a \gam \gam} = g_{\gam}\left(\alpha m_{a}/\pi\Lambda_{\te{QCD}}^{2}\right)$.\\
\indent The remaining terms in Eq.~\ref{Axion Power} contain the properties of the cavity detector and other experimental apparatus: $\beta$ describes the cavity's coupling to the antenna responsible for extracting power out of the cavity, $B$ is the external magnetic field the cavity sits in, $C$ is a cavity-mode-dependent geometric factor that describes the overlap between the mode's electric field and the applied magnetic field, $V$ is the cavity volume, $Q_{0}$ is the internal quality factor of the cavity, $Q_{L}$ is the loaded quality factor, and $\Delta f_{c} = f_{c} - f_{a}$ describes the detuning between the axion frequency $f_{a} = m_{a}c^{2} / h$ and the cavity frequency $f_{c}$.\\
\indent The scan speed of a cavity haloscope depends on the readout scheme used. If a photon-counting device is used, the scan speed scales with the cavity parameters as $df/dt \propto (CV)^{2}Q_{0}^{2}$. If instead an axion experiment uses a linear amplifier, the scan speed scales with the cavity parameters as $df/dt \propto (CV)^{2}Q_{0}$. We derive these different scaling rates in Appendix~\appScanSpeed. \\
\indent We note that the addition of dielectrics such as sapphire in axion haloscopes is typically associated with a loss in sensitivity \cite{baiUseDielectricElements2023}. This is because dielectric cavities typically trade an improved quality factor for a reduction in $CV$. For a search utilizing a low noise amplifier, this typically results in a net loss in scan speed given the stronger dependence on $(CV)^{2}$ compared to the quality factor $Q_{0}$. However, we are interested in pairing this cavity not with a quantum-limited amplifier, but rather with a photon counting device \cite{dixitSearchingDarkMatter2021a, braggioQuantumEnhancedSensingAxion2025}. Such photon-counters typically involve a buffer resonator that needs a sufficiently narrow bandwidth to resolve the Stark shift of the resonator-qubit system, which is typically on the order of $\mathcal{O}(1\, \mathrm{MHz})$. Impedance-matching to such a buffer requires a high-$Q$ cavity like the one described here. In addition, this cavity design gives a favorable scan speed; since $Q_{0}$ scales the same way as $CV$, it is easier to offset the loss in effective cavity volume by improving $Q_{0}$.
\section{\label{sec:Exp Setup}Experimental Setup}
\begin{figure}[htbp]
  \centering
  \begin{minipage}{0.48\columnwidth}
    \centering
    \begin{minipage}[b][0.22\textheight][b]{\linewidth}
      \centering
      \includegraphics[width=\linewidth, height=0.22\textheight, keepaspectratio]{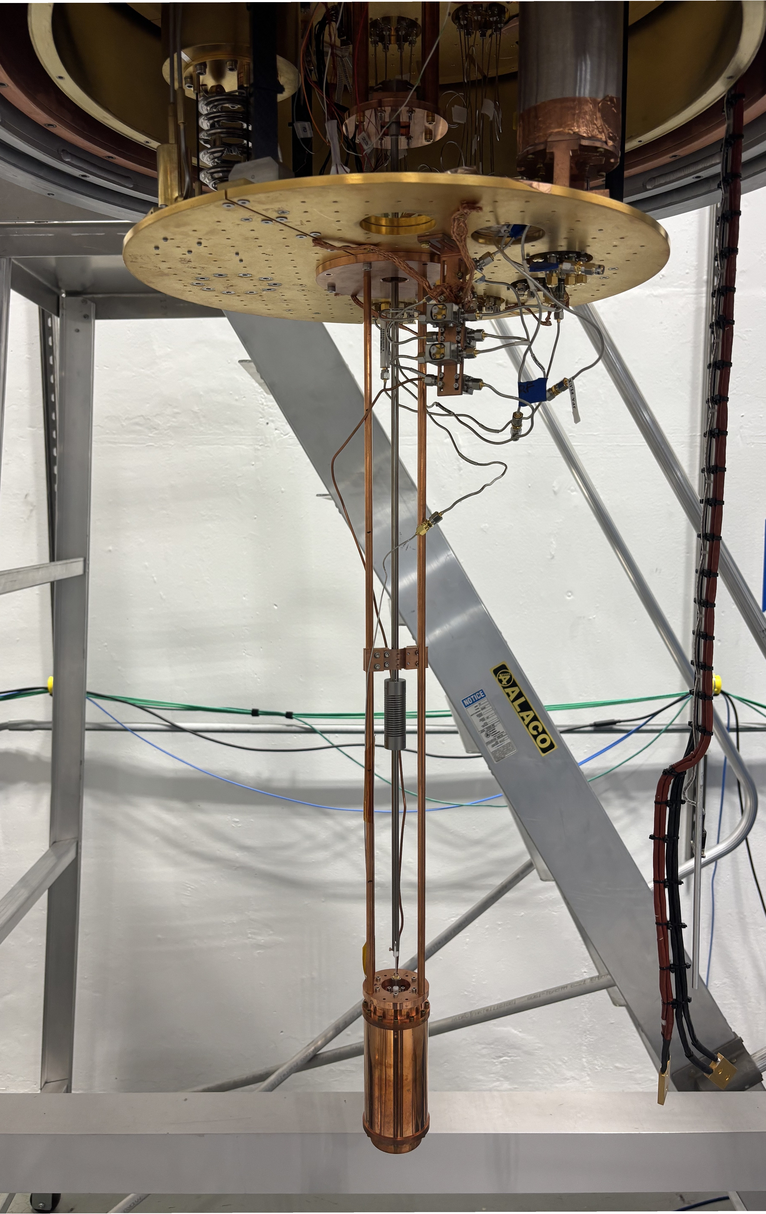}
    \end{minipage}\\(a)
  \end{minipage}\hfill
  \begin{minipage}{0.48\columnwidth}
    \centering
    \begin{minipage}[b][0.22\textheight][c]{\linewidth}
      \centering
      \includegraphics[width=\linewidth, height=0.22\textheight, keepaspectratio]{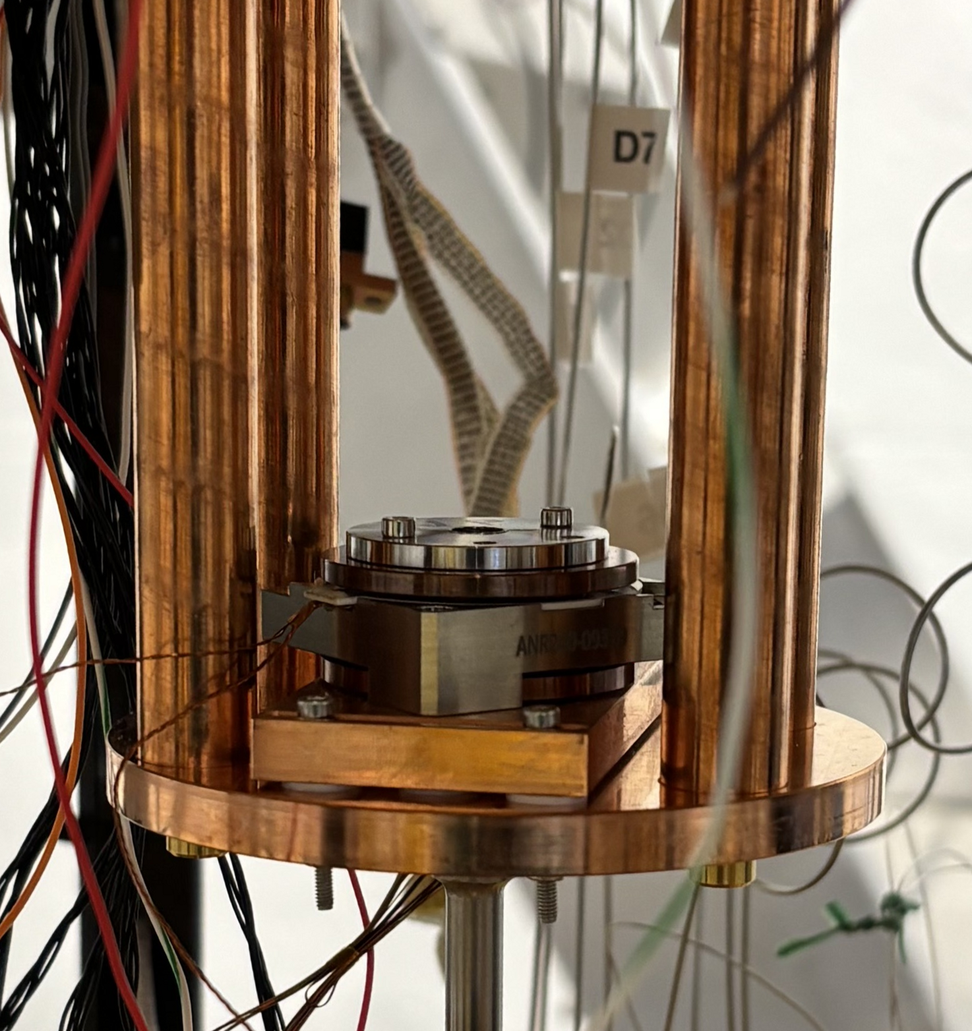}
    \end{minipage}\\(b)
  \end{minipage}

  \vspace{1mm}

  \begin{minipage}{0.48\columnwidth}
    \centering
    \begin{minipage}[b][0.172\textheight][b]{\linewidth}
      \centering
      \includegraphics[width=\linewidth, height=0.172\textheight, keepaspectratio]{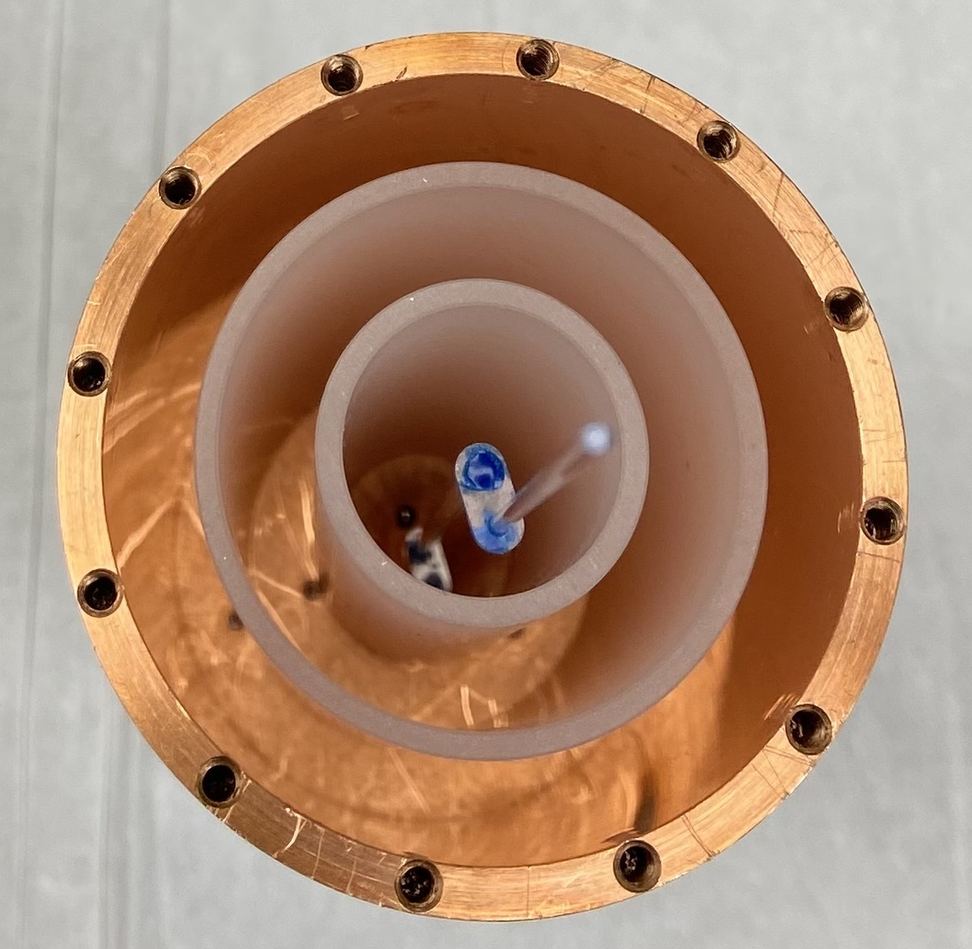}
    \end{minipage}\\(c)
  \end{minipage}\hfill
  \begin{minipage}{0.48\columnwidth}
    \centering
    \begin{minipage}[b][0.172\textheight][b]{\linewidth}
      \centering
      \includegraphics[width=\linewidth, height=0.172\textheight, keepaspectratio]{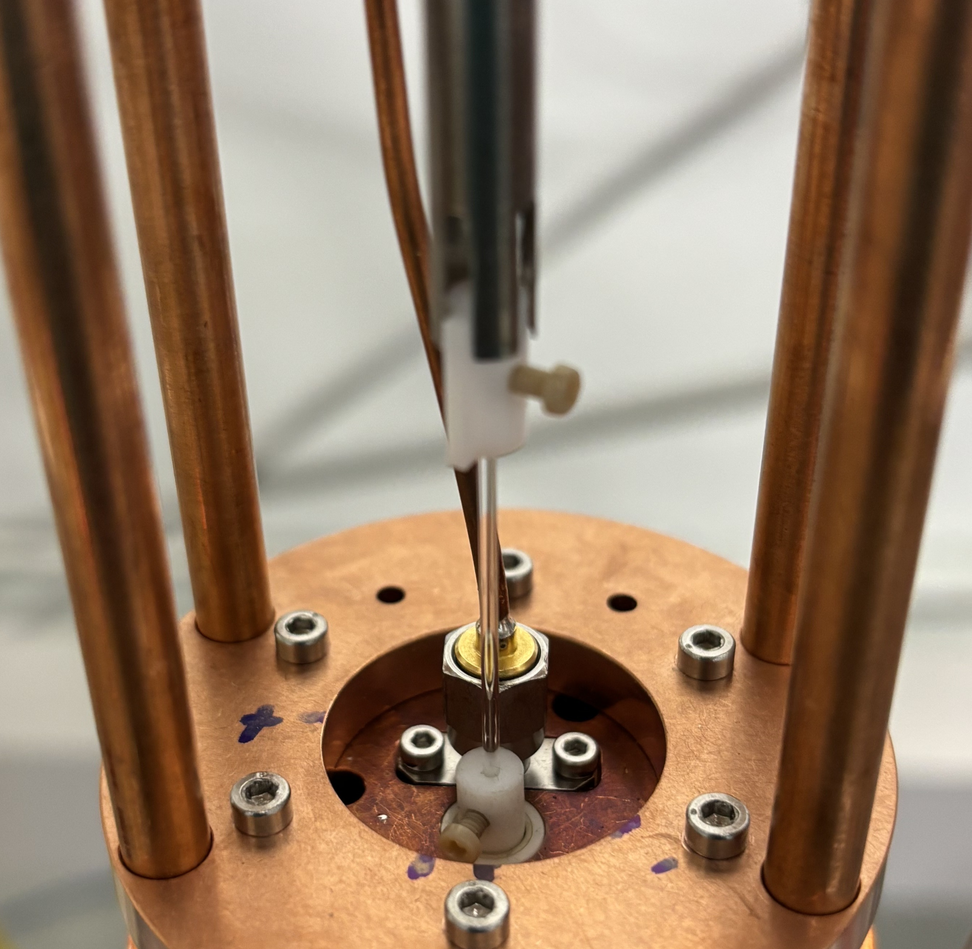}
    \end{minipage}\\(d)
  \end{minipage}
  \caption[Cavity Assembly]{Experimental setup. (a) Haloscope and piezo motor mounted in the dilution refrigerator. (b) Piezoelectric motor mounted to the $1$ K plate of the refrigerator. Teflon washers thermally isolate the motor plate during a cooldown to prevent atmosphere from freezing on the motor. (c) Cavity haloscope inner volume. Concentric sapphire shells confine the mode away from the copper walls while a sapphire tuning rod rotates in a circular path to tune the mode. (d) Tuning rod coupling to the piezo motor. A stainless steel rod runs from the motor to the cavity, and a Teflon gear couples the tuning rod to the stainless steel rod.}
\label{fig:Cavity Setup}
\end{figure}
\indent The cavity haloscope used in this experiment is based on a previously published design \cite{divora2022highqmicrowavedielectricresonator}, our version of which is shown in Figure~\ref{fig:Cavity Setup}. It is composed of an OFHC copper cylinder with radius $r = 25.82$ mm and height $h = 10$ cm. Two concentric sapphire shells, with inner radii $r_{1} = 9.55$ mm and $r_{2} = 17.72$ mm and thickness $w = 1.87$ mm, confine a $\te{TM}_{010}$-like mode inside the inner volume to reduce losses on the copper barrel and improve the quality factor. While the chosen mode for axion conversion resembles a $\te{TM}_{010}$ mode inside the first sapphire shell, in reality this is the higher order $\te{TM}_{030}$ mode with its outer lobes suppressed by the Bragg metamaterial cylindrical walls.  This cavity design gains higher $Q$ at a cost of the effective volume. To further improve the quality factor, a $10$ mm tall taper with half-angle $\theta = 40.27$ degrees is introduced to the endcaps over the inner shell volume to reduce losses in the copper. A similar experiment using the $\te{TM}_{030}$ mode used a clamshell design to achieve a limited tuning range of 58~MHz~ \cite{sardoinfirriSearchPostinflationaryQCD2025}. By keeping the cavity sealed and instead using a sapphire tuning rod, we achieve a much larger tuning range of 1.6~GHz.  Traditionally, a copper or metallic tuning rod has been the standard tuning mechanism for axion haloscopes \cite{rapidisCharacterizationHAYSTACAxion2019, bartramAxionDarkMatter2021}, but these couple $\te{TEM}$ modes into the cavity that can hybridize with our tuning mode and contribute more copper surface area over which to incur loss. In contrast, sapphire is a low-loss material. While it does not couple $\te{TEM}$ modes into the cavity, it does couple higher-order modes to be supported purely in the sapphire rod itself. In other words, the presence of sapphire increases the density of modes contained between a given frequency interval. The rod itself is a cylinder with radius $r_{r} = 1.05$ mm, making the ratio of the tuning rod radius to the inner radius of the first sapphire shell $r_{r}/r_{1} \approx 1/10$, a value that is thought to optimize the tuning range and scan speed for a dielectric tuning rod \cite{baiUseDielectricElements2023}. \\
\begin{figure}[htbp]
\centering
\includegraphics[width=\columnwidth, height=0.3\textheight, keepaspectratio]{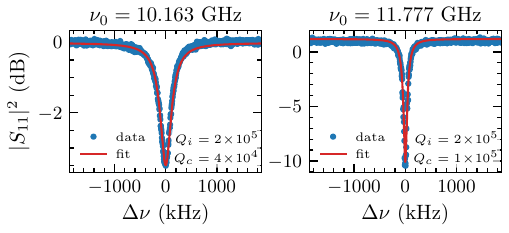}
  \caption[Full Cavity Tuning Range]{Reflection spectra of the haloscope cavity at both ends of its full tuning range with fit parameters. This data was taken with two different antenna couplings. The full tuning range of the cavity is from $10.16-11.77$ GHz, about $16\%$ tunability. The full tuning range is achieved when the rod is rotated over $180$ degrees of a circular path that is off-center relative to the cavity, moving the tuning rod from its closest point to the cavity center to the farthest point.}
  \label{fig:full tuning}
\end{figure}
\indent The tuning rod can be seen in Figure~\ref{fig:Cavity Setup}(c). Two other sapphire pieces run perpendicular to this tuning rod to couple the tuning rod to a sapphire axle driven by the piezoelectric motor. This axle rotates in place, and the tuning rod traces a circle inside the cavity volume. The blue material seen in the Figure is the cryogenic epoxy used to join the multiple pieces into one unit. Note the axle's axis of rotation is off-center relative to the cavity. Given the symmetry of the cavity, only $180$ degrees of the rod's rotation corresponds to a unique tuning range. The full tuning range is over $10.16-11.77$ GHz and can be seen in Figure \ref{fig:full tuning}. These two points correspond to when the tuning rod is closest to the cavity center and when the tuning rod is furthest away from the cavity center. For the experiment below, typical cavity parameters were $Q_{0} \approx 5 \times 10^{5}$ and $\beta \approx 14$, well into the overcoupled regime, with typical loaded quality factor $Q_{L} \approx 3.7 \times 10^{4}$. While not optimal from a scan-speed perspective, this coupling was chosen to reduce the number of tuning steps needed to cover a moderately wide frequency band for this demonstration, while still preserving sensitivity to new axion parameter space.\\
\indent We drive the tuning rod with a piezoelectric motor (Attocube ANR240, non-magnetic), shown in Figure~\ref{fig:Cavity Setup}(b). The motor was tested by the vendor in a better vacuum environment than we achieve in our refrigerator \cite{attocubePrivate2026}, so we could not rule out the possibility of residual nitrogen, oxygen, or water freezing on the motor and causing it to jam. We use Teflon washers to thermally isolate the piezo motor from the refrigerator plate it is heat sunk to, causing any residual gas to freeze elsewhere rather than on the motor gears. The tradeoff is that the motor takes longer to dissipate heat at cryogenic temperatures. We find even a few degrees difference between the motor and the refrigerator during a cooldown has been sufficient for reliable motor operation. This also kept the cooling time at cryogenic temperatures on the order of one minute for a typical tuning step in the experiment below. \\
\indent The motor is connected to a hollow stainless steel (SS) shaft that runs the roughly $70$ cm from the motor to the sapphire axle. The material, length and shape keeps the heat flow from the motor to a minimum. To dampen mechanical vibrations which may travel down the SS shaft, beryllium-copper fins were attached to the cavity support mechanism and pressed lightly against the rod. To accommodate possible geometric imperfections, a stainless-steel cryogenic U-joint is incorporated into the shaft to allow torque to be delivered even with some misalignment. The fins and U-joint are shown in Figure~\ref{fig:Cavity Setup}(a). To keep the sapphire rod cold, it is coupled to the SS shaft and to cryogenic bearings using adapter pieces made of teflon, shown in Fig.~\ref{fig:Cavity Setup}(d). \\
\indent Some experiments report difficulty cooling their tuning rods to the same temperature as their cavity. We do not have a direct measurement of the temperature of our rod, but we can bound the rod temperature relative to our noise floor. If the rod were at a higher temperature than the noise floor we would observe a cavity-bandwidth-wide excess in any power spectrum measurement. We did not see such a feature when operating our cavity with a Josephson Parametric Amplifier (JPA). If our HEMT noise matches the datasheet then this JPA had a noise temperature of roughly $2$ K corresponding to 4~photons at 10~GHz. Future measurements of this rod temperature are needed to constrain this number further as the experiment aims to be coupled to photon counters sensitive to noise below the standard quantum limit.
\indent Finite-element simulations are used to determine the form factor of this cavity with the tuning rod present. We simulate the bare form factor over the majority of the tuning range, with no reduction from avoided crossings, to be $C = 0.015$. We do notice the form factor is larger $C \approx 0.026$ when the tuning rod is furthest away from the center, but that it smoothly reduces to $C = 0.015$ as the rod begins moving towards the center. We discuss this further in Appendix \appFormFactor. This form factor is defined relative to the full volume of the cavity, which is why this value is much smaller than the ideal $C = 0.69$ for a conventional copper cavity's $\te{TM}_{010}$ mode. Given that the form factor appears in Equations \ref{Axion Power} and \ref{scan speed} in the combination $CV$, we find the more useful comparison to be the dimensionless ratio $CV/\lambda^{3}$. This metric for our cavity near $10$ GHz is $CV/\lambda^{3} \approx 0.12$ compared to $CV/\lambda^{3} \approx 1$ for a bare copper cavity at the same frequency. This factor of roughly $8$ reduction in the effective cavity volume is the penalty for having the sapphire shells. But we will show that with a photon counter this penalty is compensated for by the $25\times$ improvement in the measured quality factor this cavity has over an equivalent bare copper cavity. \\
\indent The remainder of the experimental setup is described in the Supplemental Material. A dilution refrigerator provides the $20$ mK environment for the cavity, with a $14$ Tesla superconducting solenoid magnet attached to the $4$ Kelvin stage of the dilution refrigerator. For this experiment, the magnet was only ramped to $10$ T. Instabilities in the lab's cryogenic operations precluded the use of the JPA and so a HEMT amplifier with 4~K noise temperature provided the noise floor for the axion search experiment.
\section{\label{sec:Data Collection}Data Collection and Processing}
\begin{figure}[htbp]
\centering
\includegraphics[width=\columnwidth]{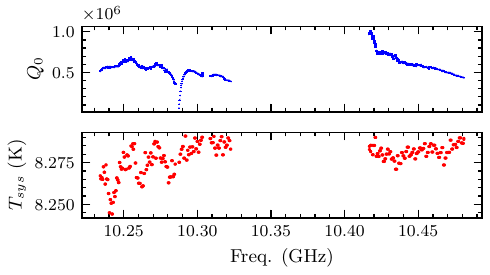}
  \caption[Cavity Q and System Noise Temperature]{System parameters for every tuning step remaining after data cuts. Top plot: Unloaded quality factor $Q_{0}$ plotted as a function of cavity frequency $f_{c}$. Bottom plot: System noise temperature, assuming the full HEMT gain and added noise given by the datasheet.}
  \label{fig:Qs and Tsys}
\end{figure}

\indent To demonstrate this cavity's use as a haloscope, we tuned the cavity and collected data with the magnet ramped to $10$ T over the frequency range $10.25 - 10.45$ GHz, roughly $12\%$ of the entire tuning range. A typical tuning step proceeded as follows. First, the cavity was tuned to its new frequency, followed by characterization measurements of the fridge temperatures and reflection measurements of the cavity using a vector network analyzer (VNA). From those measurements we perform a preliminary check for avoided crossings by verifying the phase swing across the resonator remained above $300$ degrees, which is expected for an overcoupled cavity. If this check failed we manually tuned the cavity through the crossing before resuming data collection. After this, a local oscillator downconverted the cavity frequency to $13.5$ MHz, and the signal was digitized and averaged $10^{4}$ times over $10$ seconds. Digitization was implemented with the AMD ZCU111 RFSoC running custom firmware written by the Fermilab QICK team \cite{stefanazziQICKQuantumInstrumentation2022}. Each tuning step took just under one minute, leading to an experimental duty cycle of $18\%$. After all of the data was collected, additional data cuts were made to ensure the error on the quality factor stayed below $10\%$ and the error on the form factor below $30\%$. More details can be found in Appendix \appUncertainty.\\
\indent The data that remained after these cuts is shown in Figure~\ref{fig:Qs and Tsys}. The large gap in data is due to data cuts caused by multiple avoided crossings in the region. Once all of the crossing modes that led to the large gaps in Figure~\ref{fig:Qs and Tsys} have been identified in simulation, the degradation to the form factor can be modeled and data closer to the avoided crossings can be kept.  Future experiments could also use cavities of different lengths to shift the frequencies of the non-tuning modes to cover the frequency coverage gaps associated with the current cavity. 

\begin{figure}[tp]
\centering
\includegraphics[width=0.85\columnwidth]{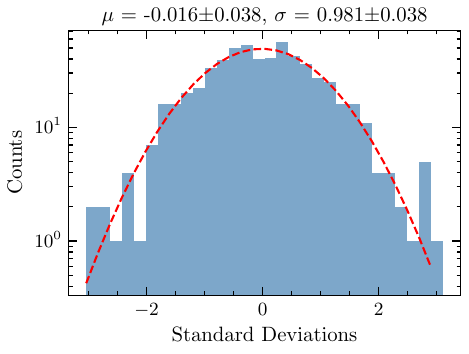}
  \caption[Data Histogram]{Example spectrum before convolution with a filter matched to the axion lineshape. The data is described by Gaussian statistics.}
  \label{fig:Data Hists}
\end{figure}

\begin{figure}[htbp]
\centering
\includegraphics[width=\columnwidth]{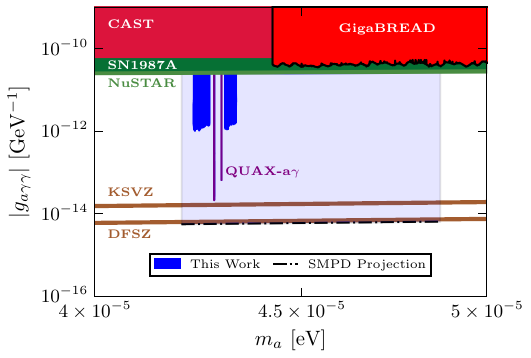}
  \caption[Axion Exclusion]{The exclusion limit from this work at $90\%$ confidence, highlighting the KSVZ and DFSZ model targets. Also shown are previous results from the QUAX$-a\gamma$ experiment in Padova, Italy \cite{sardoinfirriSearchPostinflationaryQCD2025, alesiniSearchInvisibleAxion2021}, GigaBREAD \cite{hoshinoFirstAxionlikeParticle2025}, CAST \cite{castcollaborationNewUpperLimit2024}, SN1987A \cite{manzariSupernovaAxionsConvert2024}, and NuSTAR\cite{ruzNuSTARAxionHelioscope2025}. The dashed line indicates projected $3 \sigma$ sensitivity assuming this cavity, $Q_{0} = 5 \times 10^{5}$, $CV = 3.4\, \mathrm{cm}^{3}$, with more optimal coupling $\beta = 1$, in a $14$ T magnetic field, read out with a single microwave photon detector (SMPD) with demonstrated parameters $n_{\te{th}} = 10^{-5}$ and quantum efficiency $\eta = 0.25$ \cite{dixitSearchingDarkMatter2021a, mayNoiseMitigationSingle2025}. With the prototype cavity, the scan would take $\sim 250$ years, but modest improvements ($V\rightarrow 2V$, $Q_0\rightarrow 4 Q_0$) would reduce the integration budget to $\sim 4$~years. These numbers are a large improvement over the 400,000~year scan time required for a conventional copper cavity.  This plot was uses limits compiled in Ref.~\cite{ohareCajohareAxionLimitsAxionLimits2020}.}
  \label{fig:Axion Exclusion}
\end{figure}

\indent After data cuts, the $531$ remaining tuning steps were each treated as an independent experiment, rather than combined into a single representative spectrum. Following a standard procedure in the field \cite{brubakerHAYSTACAxionSearch2017, bartramAxionDarkMatter2021}, each tuning step's spectrum was then normalized to its baseline using a Savitzky-Golay polynomial fit.  From this normalized spectrum we subtracted $1$ to obtain a spectrum of positive or negative fractional power excesses or deficits relative to its baseline. We then convolved the data with a filter matched to the axion lineshape. After this matched filter, any axion signal would have been coherently summed while any noise would add in quadrature, giving a net SNR improvement for any underlying signal. 
\indent The next step is to define a test statistic from the individual frequency bin data points $X_{i}$ through which we can gauge discovery or exclusion. Following literature in the high-energy field \cite{readModifiedFrequentistAnalysis2000,readPresentationSearchResults2002, junkSearchesLEP2001}, we define the test statistic $Q$ as
\begin{equation}\label{Test Statistic}
  Q = \f{\Phi(X_{i} - \mu)}{\Phi(X_{i})}
\end{equation}
where $\Phi$ is the standard normal cumulative distribution function, assuming that the data has been normalized to its standard deviation so that $\sigma = 1$. $\mu$ represents the signal strength of the axion signal in units of standard deviation. The advantage of this test statistic is that it protects against claims of arbitrary precision in the event of a large downward fluctuation in $X_{i}$. In order to claim a discovery, a measured power excess $X_{i} \geq 5\sigma$ is required. We did not observe any $5\sigma $ events, in which case we proceed with setting a limit on the axion coupling to photons for that frequency bin. We set a $90\%$ confidence level exclusion, by solving Equation \ref{Test Statistic} for the signal strength $\mu$ such that $Q \leq 1-0.9 = 0.1$. This means the found value of the signal strength $\mu$ and all stronger signals are excluded with $90\%$ confidence, as shown in Figure~\ref{fig:Axion Exclusion}. \\
\indent The uncertainty on the reported values of $|g_{a\gam \gam}|$ is discussed in depth in Appendix \appUncertainty. The largest uncertainties come from the system noise temperature in the absence of a direct measurement of our HEMT gain, and from the form factor in the presence of avoided crossings. These contribute a total of $20\%$ uncertainty on the axion coupling. 
\section{\label{sec:Conclusion}Conclusion}
A widely tunable cavity is a basic need for any axion haloscope experiment, and the cavity used in this work achieves a tuning bandwidth comparable to that of dish antenna experiments \cite{hoshinoFirstAxionlikeParticle2025} while maintaining the favorable resonant enhancement of a microwave cavity. 
\indent However, with the cavity lifetimes presented here with $\beta = 2$ (our run was over-coupled at $\beta \approx 14$) and operating a quantum-limited experiment, we estimate it will still take about $10^{9}$ seconds for this cavity to reach DFSZ coupling with $3\sigma$ sensitivity for a single tuning step, and hence roughly $1,000,000$ years to cover the entire tuning range in axion mass with a linear amplifier. Similarly, since the scan speed for a linear amplifier scales as $df/dt \propto (CV)^{2}Q_{0}$, the factor of $\sim 8$ degradation to $CV$ relative to a bare copper cavity offsets the factor of $25$ improvement to the quality factor of a bare cavity. Thus this cavity has no advantage compared to a bare copper cavity for experiments with linear amplifiers. If instead we used a photon counter, then $df/dt \propto (CV)^{2}Q_{0}^{2}$ scaling means this cavity maintains a factor of $\sim 9$ improvement in scan speed relative to a bare copper cavity. In addition to the scan speed argument, photon counters have already demonstrated thermal occupation numbers much lower than the quantum limited $n_{\te{th}} = 1$, with some groups reporting $n_{\te{th}} \approx 10^{-5}$ \cite{mayNoiseMitigationSingle2025} for example. As seen in Appendix C, this cavity could reach the DFSZ benchmark model across its entire tuning range in roughly $250$ years with a photon counter, assuming comparable photon counters to those already demonstrated in Refs.~\cite{dixitSearchingDarkMatter2021a, mayNoiseMitigationSingle2025} and using a $14$ T magnetic field. A future direction will be to lengthen the cavity by a factor 2, following Ref. \cite{divora2022highqmicrowavedielectricresonator}, improving both the volume $V$ and the quality factor $Q_0$ by reducing the impact of the bare copper endcaps.  The $250$ year integration time would then be reduced to $4$ years or less, laying the groundwork for a definitive DFSZ search between $10-12$ GHz.
\section{Acknowledgments}
\begin{acknowledgments}
We want to acknowledge R. Di Vora for helpful conversations about photonic bandgap cavities, the QICK team at Fermilab \cite{stefanazziQICKQuantumInstrumentation2022}, namely Horacio Arnaldi and Diego Martin for writing our digitization firmware, Daniel Bowring and Wenjie Yao for assistance with cavity simulations, Don Mitchell and Parth Gandhi for engineering advice, and Elizabeth Field for help with data collection. This work was produced by Fermi Forward Discovery Group, LLC under Contract No. 89243024CSC000002 with the U.S. Department of Energy, Office of Science, Office of High Energy Physics, and supported by its QuantISED program. Publisher acknowledges the U.S. Government license to provide public access under the DOE Public Access Plan  (http://energy.gov/downloads/doe-public-access-plan).
\end{acknowledgments}

\newcounter{myappendix}
\renewcommand{\themyappendix}{\Alph{myappendix}}

\appendix
\appendix
\section{Appendix \appFormFactor: Form Factor Simulations}
\begin{figure}[htbp]
\centering
\includegraphics[width=\columnwidth]{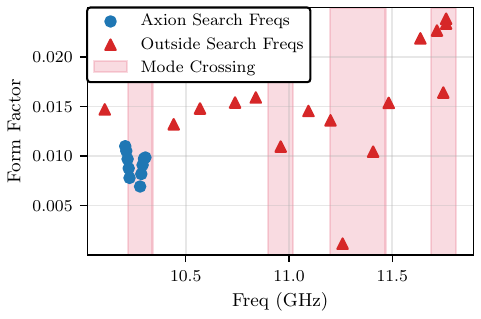}
  \caption[Simulated Form Factor]{Simulated form factor of the $\te{TM}_{030}$ mode for select points across the entire cavity tuning range. Blue points correspond to points within the frequency band of the axion search described in this paper. The cluster of points above $11.5$ GHz correspond to when the tuning rod is furthest from the cavity center and perturbs the original field profile the least. A few mode crossings are observed and highlighted in red. If the crossing mode has negligible form factor, then we expect the form factor to drop to $C_{030}/2$ at maximal mixing.}
  \label{fig:Sim FF}
\end{figure}
\indent Without a measurement of the local field profile inside the cavity, we must rely on simulation to understand the form factor of a cavity. Figure \ref{fig:Sim FF} shows the form factor of the $\te{TM}_{030}$ mode across the tuning range. \\
\indent There are two independent factors that go into the form factor values above. First is the geometric factor due to the overlap between the $\te{TM}_{030}$ mode and the external magnetic field in the absence of any mode crossings. In general this geometric factor is a function of the tuning rod position inside the cavity but typically the form factor is assumed to be constant throughout the tuning region. However, Figure \ref{fig:Sim FF} does exhibit some dependence on the tuning rod position; the form factor is larger, $C_{030} \approx 0.026$, when the tuning rod is furthest away from the cavity, and then drops to $C_{030} \approx 0.015$ as the tuning rod moves towards the center of the cavity and the $\te{TM}_{030}$ mode crosses $\sim 11.5$ GHz. For the search above where the entire frequency range is well below $11.5$ GHz, we can safely assume we are in the $C_{030} = 0.015$ regime. \\
\indent The next factor that plays a role in the simulated form factor is the presence of avoided crossings. If the crossing mode has negligible form factor then as a function of the mixing angle $\theta$ we expect the form factor to degrade as $C_{030}(\theta) \propto \cos^{2}(\theta)\, C_{030}$ so that at maximal mixing the form factor has been reduced to $C_{030}/2$. However if the mixing mode has appreciable form factor it is possible to degrade the original form factor beyond this factor of $2$ reduction. While this is unlikely, it cannot be ruled out without identifying the crossing modes, and the reduced bare form factor $C_{030} = 0.015$ makes this cavity more susceptible to non-negligible crossing-mode form factors. For this reason we chose to avoid approaching any mode crossings. 
\section{Appendix \appUncertainty: Uncertainty}
\begin{figure}[htbp]
\centering
\includegraphics[width=\columnwidth]{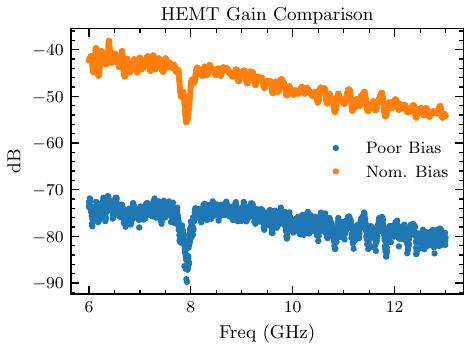}
  \caption[HEMT Gain Bound]{Transfer function through our HEMT amplifier at cryogenic temperatures at two different biases. Without a direct measurement of the HEMT gain, a bound can be placed on the HEMT gain by biasing away from the nominal bias corresponding to maximum gain and watching the gain reduce until it vanishes beneath the next-limiting noise floor. This measurement confirms the HEMT gain to be at least $28$ dB over the cavity tuning band.}
  \label{fig:HEMT Gain}
\end{figure}

\begin{table}[htbp]
\centering
\caption[Uncertainty Breakdown]{Fractional uncertainties on relevant parameters, in order of their contribution to the total uncertainty on the axion coupling. The dominant uncertainties come from the system noise temperature $T_{sys}$ and the form factor $C_{030}$. The uncertainty on $T_{sys}$ can be improved with dedicated measurements of the HEMT noise and gain, while the uncertainty on the form factor can be improved with more simulations run over the entire tuning range to fill in the gaps of Figure \ref{fig:Sim FF}.}
\label{tab:Uncertanties}
\begin{ruledtabular}
\begin{tabular}{cc}
Parameter & Fractional Uncertainty ($\%$) \\
\colrule
 $C_{030}$ & $30\%$ \\
 $T_{sys}$ & $25\%$ \\
 $Q_{L}$ & $1.5\%$ \\
 $\f{\beta}{1+\beta}$ & $\leq 1\%$ \\
 $\eta_{L}$ & $\leq 1\%$ \\
 $B$ & $\leq 1\%$ \\
 \colrule
 $|g_{a \gam\gam}|$ & $20\%$
\end{tabular}
\end{ruledtabular}
\end{table}
\indent The dominant sources of uncertainty are the following:
\begin{equation}
    |g_{a \gam \gam}| \propto (\f{T_{sys}}{\eta_{L} (\f{\beta}{1+\beta})C_{030}VQ_{L}})^{\f{1}{2}}.
\end{equation}
where the only parameter not already introduced is the attenuation between the cavity and the first-stage amplifier, $\eta_{L} \approx 8-10$ dB for this setup. The magnetic field is not listed in this equation because it is measured to better than $1\%$, but it will be included in the final table. These then contribute a fractional uncertainty on the coupling of the form:
\begin{equation}
    \f{\delta |g_{a\gam\gam}|}{|g_{a\gam\gam}|} \propto \sqrt{(\f{1}{2}\f{\delta T_{sys}}{T_{sys}})^{2} + (\f{1}{2}\f{\delta C_{030}}{C_{030}})^{2} + (\f{1}{2}\f{\delta \eta_{L}}{\eta_{L}})^{2} + ...}.
\end{equation}
\indent The reported uncertainties on the couplings are presented in Table \ref{tab:Uncertanties}, with the leading uncertainties coming from the system noise temperature and the form factor. For this setup, the dominant noise source is the HEMT added noise, which contributes to the overall system noise temperature as:
\begin{equation}
        T_{sys} = T_{4K} + T_{H} + \f{T_{RT} + T_{RT1}}{G_{H}} + \f{T_{RT} + T_{RT2}}{G_{H}G_{RT1}} + ...
\end{equation}
where $T_{4K}$ is the ambient $4$ Kelvin noise on the HEMT line, $T_{RT}$ the same for the room temperature thermal noise, the subscripts RT1 and RT2 denoting the first and second room temperature amplifiers, and $T_{H}$ denoting the added noise from the HEMT. Note that the total noise added by every amplifier is equal to its own added noise as well as the ambient thermal noise on the line as seen by that particular amplifier. We used $T_{RT} = 300$ K to be conservative. In the absence of a dedicated measurement of the HEMT noise we used the value from the datasheet $\sim 3.7$ K and do not report any error on this number. The other term that plays a role in this equation is the HEMT gain $G_{H}$. Similar to the HEMT noise the full HEMT gain cannot be measured without a dedicated setup, but we can bound $G_{H}$ with our existing setup. This is done by biasing the HEMT to its maximum gain as measured with a VNA, and then biasing the HEMT away from the optimal settings to observe the gain decrease until the signal falls below the next-limiting noise floor. The result of such a measurement is seen in Figure \ref{fig:HEMT Gain}, which bounds the HEMT gain to at least $28$ dB of gain over the cavity tuning band, relative to the full $37$ dB claimed by the datasheet. Since all evidence points to the HEMT providing full gain, the datasheet gain value was assumed for the analysis. The worst-case scenario is that instead the HEMT is only providing the $28$ dB of gain we measured, in which case our error on $T_{sys}$ would be $\sim 25\%$. \\
\indent Similar to the form factor described in Appendix \appFormFactor, if the crossing mode has negligible coupling to the cavity antenna the coupling $Q_{c}$ has the same dependence on the mixing angle as the form factor $Q_{c}(\theta) \propto Q_{c}/\cos^{2}(\theta)$, causing the form factor to become a proxy for the mixing angle. The coupling mode having negligible antenna coupling is a separate question from the mode having negligible form factor. As discussed above, the inclusion of dielectrics in the cavity volume increases the mode density by supporting modes inside the dielectric. However, these modes will have no coupling to our antenna and will thus exhibit this $Q_{c}(\theta) \propto Q_{c}/\cos^{2}(\theta)$ behavior. Thus, we cut any data points whose $Q_{c}$ deviated by more than $30\%$ from its nominal value, which keeps the form factor uncertainty below $30\%$ due to mode crossings. \\
\indent The other errors come from fitting the reflection measurements to extract the quality factors and couplings, and these errors are sub-dominant to those of the form factor and system noise temperature. In total, these contribute about $20\%$ fractional uncertainty on the final reported couplings. In other words, for any of our data points in Figure \ref{fig:Axion Exclusion}, the ``true'' value of the $90\%$ CL lies in a $20\%$ band around the reported value.
\section{Appendix C: Scan Speed Derivations}
Written in terms of the internal quality factor $Q_0=(1+\beta) Q_L$ of the cavity, the predicted signal photon rate is
\begin{equation}
R_s \equiv  \eta P_\mathrm{ax} / m_a =  \frac{\eta}{2\epsilon} \frac{\beta}{(1+\beta)^2} \frac{1}{\mu_0}(g_\gamma \frac{\alpha}{\pi} \theta B)^2 C V Q_0 
\end{equation}
where $\eta$ is the quantum efficiency of the single photon detector and $Q_0 = (1+\beta) Q_L$ is the internal quality factor of the cavity. For a detection bandwidth given by the cavity linewidth, $b=f/Q_L$, and mode occupation number $n_b$, the rate of background photons is
\begin{equation}
    R_b = n_b f/Q_L.
\end{equation}
The time $dt$ needed to achieve a detection of a signal excess with 3$\sigma$ significance in Poisson statistics of photon counts is obtained from
\begin{equation}
R_s t \ge 3 \sqrt{R_s t + R_b t}
\rightarrow t \ge 9 (R_s + R_b)/R_s^2 \approx 9 R_b / R_s^2.
\end{equation}
The frequency tuning step size can be expressed in units of the loaded cavity bandwidth, e.g. for 3 steps per bandwidth, 
\begin{equation}
    \delta f = \frac{1}{3} f/Q_L. 
\end{equation}

The frequency scan speed is then
\begin{gather}
    \frac{df}{dt} = \delta f / dt = \frac{1}{27} \frac{f}{Q_L} \frac{R_s^2}{R_b} 
    = \frac{1}{27} \frac{R_s^2}{\eta n_b} \\
   \frac{df}{dt} \propto \eta^{2} \frac{\beta^2}{(1+\beta)^4} \frac{B^4 (C V)^2 Q_0^2}{n_b}. \label{scan speed}
\end{gather}
The scan speed now scales as $(C V)^2 Q_0^2$ because the factors of $Q_L$ in the frequency step size and the detection bandwidth cancel each other.  This cancellation does not occur in traditional phase-preserving linear amplification experiments, in which the detection bandwidth is instead $f/Q_a$ where $Q_a \approx 10^6$ is the predicted Doppler-broadened signal linewidth of the dark matter trapped within the gravitational potential well.  The scan speed in that case scales as the product $(C V)^2 Q_0 Q_a$ when $Q_a > Q_L$.  The phase-preserving amplifiers allow the resolution of subcomponents of the cavity bandpass of width $f/Q_a < f/Q_L$ at the cost of incurring the standard quantum limit noise, whose variance is equivalent to the Poisson noise of $n_b = 1$ photon per resolved Fourier mode.  The orders of magnitude reduction to $n_b < 10^{-4}$ of demonstrated qubit-based single microwave photon counters, however, far outweighs the small improvement in $Q_a > Q_L$ for the linear amplifiers.  Also note that $CV$ and $Q_0$ appear on equal footing in the scan speed scaling for photon counting, so it becomes more favorable to trade a reduced $CV$ for increased $Q_0$ for these experiments.   

Note that the scan rate for a given cavity is independent of frequency other than from the small dependencies hidden in $\beta$, $C$, and $Q_0$.  Also, the value $\beta=1$ for critical coupling now maximizes both the signal photon rate $R_s$ and the scan rate $df/dt$.  

Plugging in the numbers for our demonstrated cavity parameters and for demonstrated single microwave photon detectors gives
\begin{gather} \notag
    \frac{df}{dt} = 0.2 \ \mathrm{Hz/s} \times
    \left(\frac{\eta}{0.25}\right)^{2}
    \left(\frac{B}{14 \ \mathrm{T}}\right)^4 \times \\
  \times  \left(\frac{CV/\lambda^3}{0.12}\right)^2
    \left(\frac{Q_0}{5\times 10^5}\right)^2
    \left(\frac{10^{-5}}{n_b}\right)
\end{gather}
To cover the entire 1.6~GHz tuning range of the cavity with 3$\sigma$ sensitivity would therefore take 250 years. However, due to the quadratic scaling, even small improvements in $(C V Q_0)$ would drastically reduce the time required to cover this range, 1/6 of an octave in frequency.  For example, similar cavities have been demonstrated with quality factors nearing $Q_0\approx 10^7$ \cite{divora2022highqmicrowavedielectricresonator}.  Even a modest improvement to raise our measured quality factor by a factor of $4$ to $Q_{0} = 2 \times 10^{6}$ would reduce the integration time by a factor of $16$. There is also room to increase the length of the cavity and hence its volume by a factor of 2, while keeping the majority of the device in the high-field region of our magnet, thus reducing the integration time by a factor of $4$. These two cavity improvements would reduce the $250$ years to $\sim 4$ years, and further reductions may be achieved by improving quantum efficiency and reducing dark count rates of the single photon detectors. Operating 6 such cavities simultaneously could cover an entire octave in frequency in a similar total amount of experimental time.

%

\end{document}